\documentclass[twocolumn,preprintnumbers,amsmath,amssymb,superscriptaddress,floatfix]{revtex4-2}
\usepackage[dvips]{graphicx}
\usepackage{dcolumn}
\usepackage{bm}

\usepackage[dvips]{graphicx}
\usepackage{dcolumn}

\usepackage{amsmath}    
\usepackage{epstopdf}
\usepackage{graphicx,amssymb}
\usepackage{color,ulem}
\usepackage{float}
\usepackage{siunitx}
\usepackage{mathrsfs}
\usepackage{xcolor}
\usepackage{bm}

\usepackage{soul}
\usepackage{cancel}

\usepackage[%
colorlinks=true,
pdfborder={0 0 0},
citecolor=blue,
linkcolor=blue
]{hyperref}

\begin{document}
	
		\title{A blueprint for experiments exploring the Poincar\'e quantum recurrence theorem}

	\author{Bayan Karimi}
	\thanks{email:karimib@uchicago.edu;bayan.karimi@aalto.fi}
	\address{Pritzker School of Molecular Engineering, University of Chicago, Chicago IL 60637, USA}
	\affiliation{Pico group, QTF Centre of Excellence, Department of Applied Physics, Aalto University School of Science, P.O. Box 13500, 00076 Aalto, Finland}
	
	\author{Xuntao Wu}
	\address{Pritzker School of Molecular Engineering, University of Chicago, Chicago IL 60637, USA}
	
	\author{Andrew N. Cleland}
	\address{Pritzker School of Molecular Engineering, University of Chicago, Chicago IL 60637, USA}
	\address{Center for Molecular Engineering and Material Science Division, Argonne National Laboratory, Lemont IL 60439, USA}
	
	\author{Jukka P. Pekola}
	\thanks{email:jukka.pekola@aalto.fi}
	\affiliation{Pico group, QTF Centre of Excellence, Department of Applied Physics, Aalto University School of Science, P.O. Box 13500, 00076 Aalto, Finland}

	\date{\today}
	
	\begin{abstract}
	The quantum form of the Poincar\'e recurrence theorem stipulates that a system with a time-independent Hamiltonian and discrete energy levels returns arbitrarily close to its initial state in a finite time. Qubit systems, being highly isolated from their dissipative surroundings, provide a possible experimental testbed for studying this theoretical construct. Here we investigate a $N$-qubit system, weakly coupled to its environment. We present quantitative analytical and numerical results on both the revival probability and time, and demonstrate that the system indeed returns arbitrarily close to its initial state in a time exponential in the number of qubits $N$. The revival times become astronomically large for systems with just a few tens of qubits. Given the lifetimes achievable in present-day superconducting multi-qubit systems, we propose a realistic experimental test of the theory and scaling of Poincar\'e revivals. Our study of quantum recurrence provides new insight into how thermalization emerges in isolated quantum systems.
\end{abstract}

\maketitle

{\it Introduction ---} Poincar\'e recurrence, both in its classical~\cite{Poincare1890, Caratheodory1919,Kac1943,Zwanzig} and quantum~\cite{Bocchieri1957, Schulman1978,Peres1982,Peres1993,Robinett2004} forms, is relevant to discussions of thermalization, quantum chaos~\cite{Haake2018}, and the foundations of quantum thermodynamics~\cite{Campbell2025}. It places strict constraints on the long-time behavior of isolated quantum systems, in spite of their apparent equilibration over shorter time scales~\cite{Grnbaum2013, Wallace2015,Zhong2019}. According to the Poincar\'{e} recurrence theorem, an isolated system will evolve such that after a finite time, it returns arbitrarily close to its initial state. Classical recurrence theory was explored in the 19th century, whereas the quantum recurrence theory (QRT) is known only since the 1950's, from when it has been applied to time-independent quantum mechanical systems with discrete energy eigenstates~\cite{Erne2017, Chen2024}.

We can investigate QRT in coupled many-body quantum systems: consider a quantum two-level system, a qubit, that is coupled to a quantum mechanical environment with a finite number $N$ of degrees of freedom. There is a natural question of how the Poincar\'e recurrence time for the qubit scales with the  size of its environment and with the desired return fidelity. This relates intrinsically to the problem of thermalization in isolated quantum systems \cite{Mori2018, Nandkishore2015,DAlessio2016}, Loschmidt echos in quantum many-body dynamics~\cite{Karch2025}, as well as to questions of what forms a heat bath in a quantum system \cite{Pekola2024}. These issues are thought to be hard to address in experimental quantum systems, due to the difficulty in sufficiently isolating the system from its dissipative environment over the relevant time scales. Here, we provide a theoretical analysis of the Poincar\'e recurrence time scales and signature responses, with new insights into the small $N$ behavior. We then use analytical and numerical models to explore Poincar\'e recurrence in an experimentally-realizable system of $N \sim \mathcal{O}(10)$ superconducting qubits, in which signatures of Poincar\'e recurrence should be observable over reasonable time scales even in the presence of realistic levels of environmentally-induced dissipation.
\begin{figure}[h]
	\centering
	\includegraphics [width=\columnwidth] {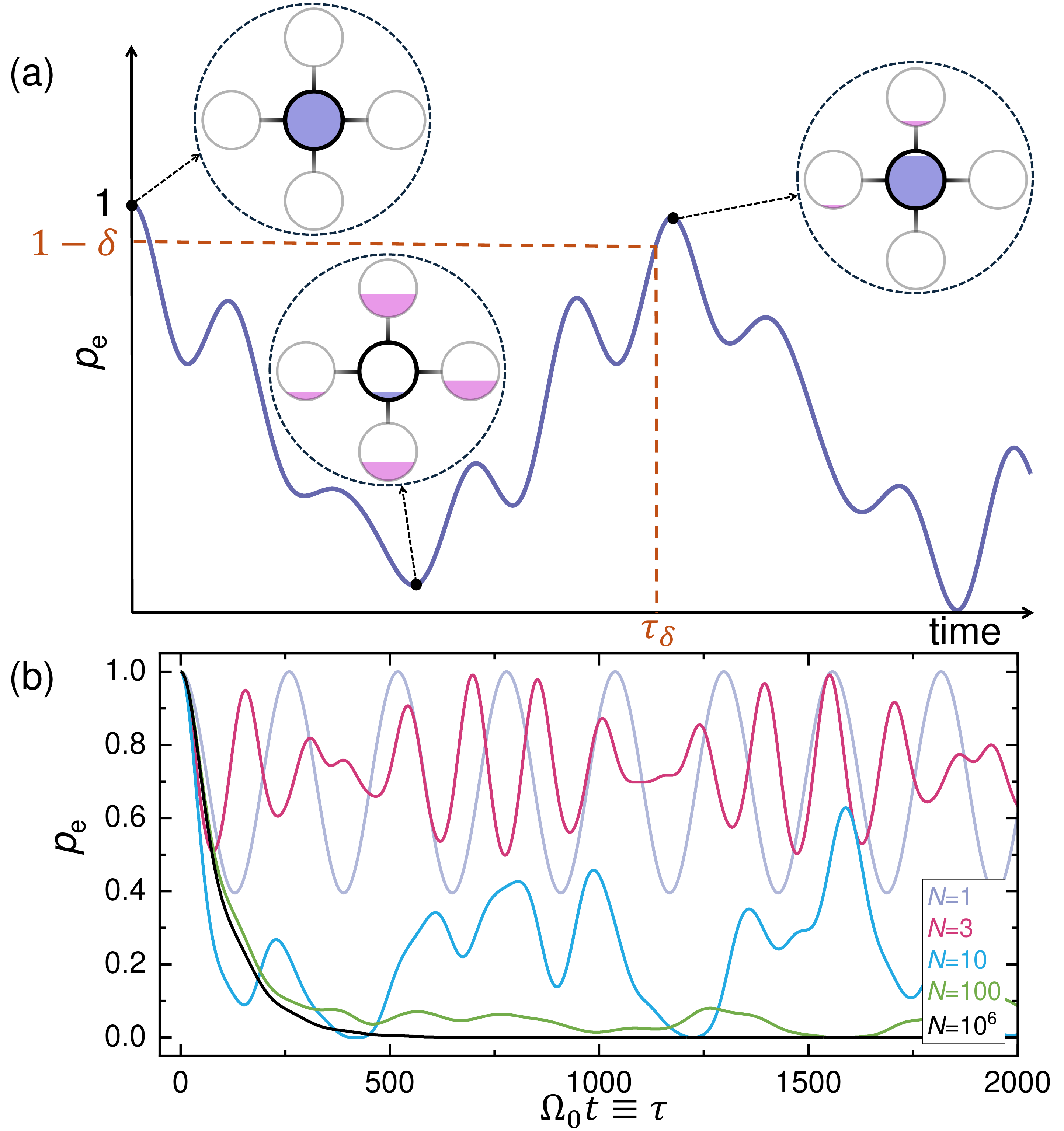}
	\caption{Concept of revivals. (a) At time $t=0$, a central ``test'' qubit is initialized in its excited state, while it is coupled to $N$  ``environmental'' qubits (here $N=4$), all of which are initially in their ground states. The solid line illustrates the resulting time evolution of the excited-state population of the central qubit. The accompanying schematic provides a conceptual visualization of the qubit configurations at various points in time. Here $\tau_\delta$ refers to the first passage time, i.e. the Poincar\'e time for a given configuration of system parameters, with $\tau_{\rm P}$ the average of $\tau_\delta$ over different configurations. (b) Examples of time traces of the excited-state population $p_e$ of the central qubit for different numbers $N$ of environmental qubits. The parameters are $\Delta\Omega/\Omega_0=0.1$, and $\Gamma_0=0.01\Omega_0$; coupling strengths are distributed uniformly from $0$ up to maximum value determined by $\Gamma_0$. For $N=1$, we see the expected sinusoidal Rabi oscillations, but with increasing $N$ the oscillations become less regular and weaker until for $N=10^6$, we see only exponential decay.
		\label{motivation}}
\end{figure}

We study a system comprising a central ``test" qubit coupled to an ensemble of $N$ ``environmental'' qubits, as shown in Fig.~\ref{motivation}\,(a), where we vary the energy splittings and coupling strengths of the qubits. We find that, as expected, the Poincar\'{e} recurrence time scales exponentially with $N$ for a given range of splittings and coupling strengths. We derive analytical results for this system in the weak linear coupling regime, and perform numerical analyses which are not restricted to the weak coupling condition. With just a few tens of environmental qubits, the system no longer recovers to its initial state on any realistic time scale. Finally, we relate our results to recently-published experimental platforms \cite{Wu2024, Andersen2025} and find that, given current relaxation and decoherence rates, experimental tests are fully feasible for multi-qubit systems with tunable couplings and energies~\cite{Kjaergaard2020,Bal2024,Tuokkola2024}. 

The main theoretical results can be obtained from the analysis of an archetypal synthetic open system described by the standard Hamiltonian $\hat{\mathcal{H}}=\hat{\mathcal{H}}_{\rm S}+\hat{\mathcal{H}}_{\rm E}+\hat{\mathcal{H}}_{\rm c}$, where $\hat{\mathcal{H}}_{\rm S} = \hbar \Omega_0 \hat{a}^\dagger \hat{a}$ represents the Hamiltonian of a central qubit with energy splitting $\hbar\Omega_0$, $\hat{\mathcal{H}}_{\rm E} = \sum_{j=1}^{N} \hbar\Omega_j \hat{b}_j^\dagger \hat{b}_j$ denotes the Hamiltonian of the environmental qubits,  composed of $N$ additional ones with energies $\hbar\Omega_j$, and $\hat{\mathcal{H}}_{\rm c} = \sum_{j=1}^{N} g_{0j}(\hat{a}^\dagger \hat{b}_j+\hat{b}_j^\dagger \hat{a})$ models the system-environment qubit interaction, assumed to be linear with a coupling strength $g_{0j}$ between the central qubit and the $j^{\rm th}$ environment qubit; $\hat{a}$ and $\hat{b}_j$ are the annihilation operators for the corresponding qubits, respectively.

We start by solving the Schr\"odinger equation, $i\hbar \partial_t |\psi_I(t)\rangle = \hat{\mathcal{H}}_{{\rm c},I}(t)|\psi_I(t)\rangle,$ for the complete system in the single excitation subspace, including the central qubit and the $N$ qubits in the environment. Here, the subscript $I$ indicates that the coupling Hamiltonian and the wave function are expressed in the interaction picture. We then obtain a set of time-evolution equations~\cite{Pekola2022} for which an analytical solution can be found by proceeding iteratively. Figure~\ref{motivation}\,(b) presents numerical examples of how the excited state population $p_e$ of the central qubit evolves when it is in contact with $N$ environmental qubits. For small $N$, we observe revivals, with individual traces demonstrating an almost periodic function~\cite{Bohr1947,Amerio1971}. For large $N$, these revivals disappear entirely.  In this figure, we parameterize the coupling by $\Gamma_0$ related to $g_{0j}$ through $\langle g_{0j}^2 \rangle = 3\Delta \Omega \Gamma_0/(2\pi \Omega_0 N)$, where $\Delta\Omega$ is the energy spread of the environmental qubits, as described below. We next build a multiscale model, where the previously isolated system is embedded in a much more complex environment that is responsible for decoherence of the qubit system. This allows analysis relevant to an actual experimental system. We include the leakage to the environment bath quantitatively, using experimentally feasible relaxation rates for the individual qubits. This multiscale model guides us as to what extent and over what time scale the $N+1$ qubits can be considered to form an isolated quantum system from the perspective of Poincar\'e recurrence. 

{\it Quantum recurrence ---} To assess quantitatively the revival process, we study a framework that can eventually be implemented experimentally. As illustrated in Fig.~\ref{motivation}\,(a), for $N=4$, we prepare the central test qubit in its excited state, and monitor its state in time as it interacts with the environment qubits, which are initialized in their ground states, a process that can be realistically implemented in an experiment using superconducting qubits~\cite{Kjaergaard2020}. 

We first demonstrate that monitoring the excited state population of the test qubit serves as a measure of quantum revivals. We define the basis formed by the single-excitation subspace $|0\rangle\equiv |1\,0\,0\,...\,0\rangle$ and $|{j}\rangle\equiv |0\,0\,0\,...1 (j^{th})\,...\,0\rangle$, where the first entry corresponds to the central qubit followed by the entries corresponding to each of the $N$ environment qubits. Assuming the arrangement shown schematically in Fig.~\ref{motivation}\,(a), at time $t=0$, the state of the system is $|\psi(0)\rangle=|1\,0\,0\,...\,0\rangle\equiv |0\rangle$. 

After a time $t$, the wave function is given by
\begin{equation}\label{wavefunction-with phase}
	|\psi(t)\rangle=\mathscr{C}_{0}(t)|0\rangle+\sum_{j=1}^{N}\mathscr{C}_{j}(t)|j\rangle, 
\end{equation} 
where $|\mathscr{C}_{0}(t)|^2$ and $|\mathscr{C}_{j}(t)|^2$ represent the population of the central qubit and the $j^{\rm th}$ one in the environment, respectively. We define the distance $d$ of the state at $t$ from the initial state as $d=\sqrt{\langle \delta\psi(t)|\delta\psi(t)\rangle}$, where $|\delta\psi(t)\rangle\equiv|\psi(t)\rangle-|\psi(0)\rangle$. We obtain $d^2=2{(1-|\mathscr{C}_{0}(t)|^2)}/({1+|\mathscr{C}_{0}(t)|})\leq 2(1-|\mathscr{C}_{0}(t)|^2)$. Thus if $1-|\mathscr{C}_{0}(t)|^2$ is small, then $d^2$ is also small, justifying the use of $p_e(t)\equiv |\mathscr{C}_{0}(t)|^2$ as a measure of the recurrence.
\begin{figure}
	\centering
	\includegraphics [width=\columnwidth] {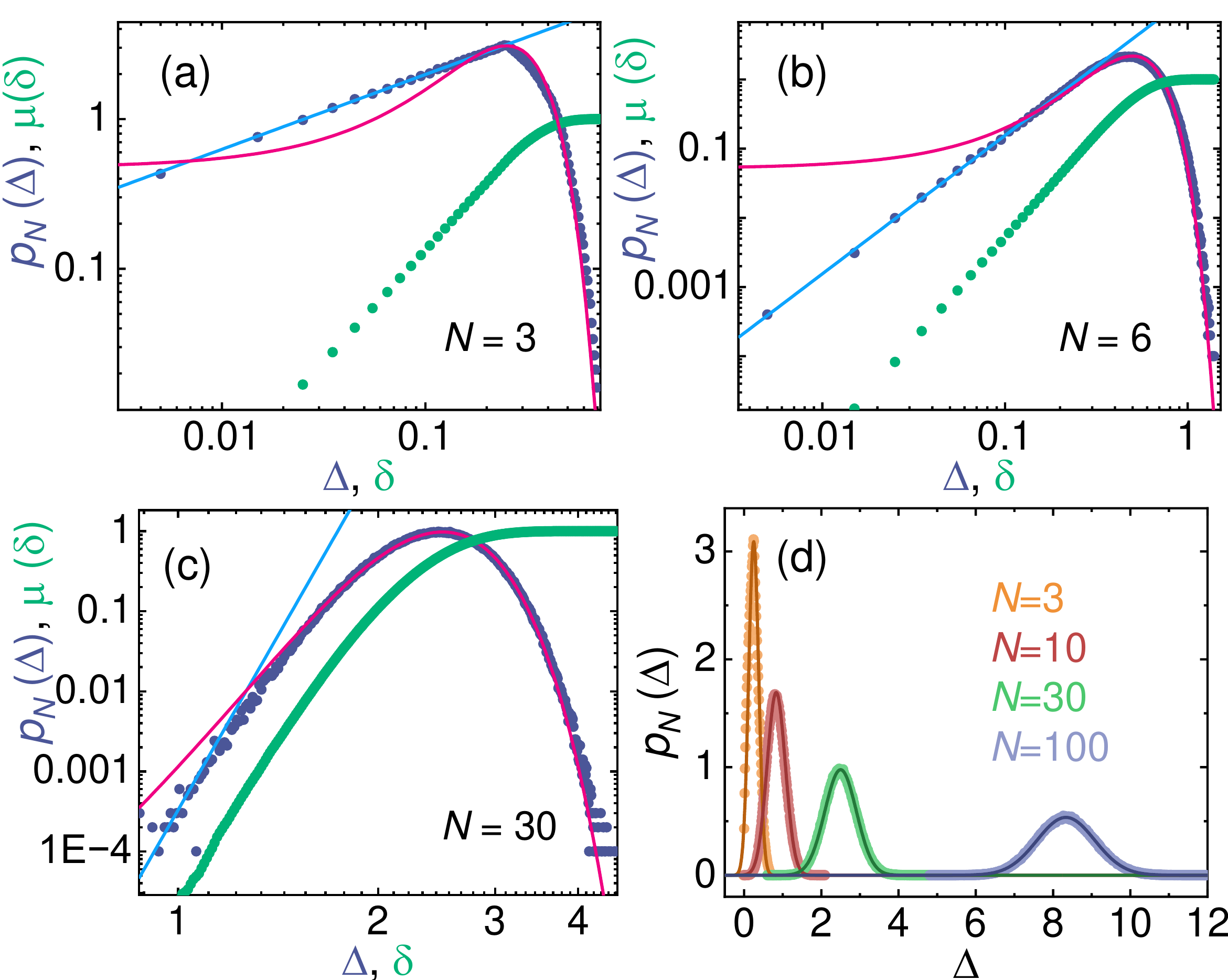}
	\caption{Probability density distributions $p_N(\Delta)$ of $\Delta\equiv \sum_{i=1}^{N}\Delta_i$ for (a) $N=3$, (b) $N=6$, and (c) $N=30$. For illustration, we set $\mathcal{T}=1$ here. The magenta line is the Gaussian distribution, the dark blue symbols show the numerical stochastic values (histogram), and the aqua line is from Eq.~\eqref{final_pNDelta1}, valid for small $\Delta$. The green symbols show the numerically calculated cumulative histogram $\mu_N(\delta)=\int_{0}^{\delta} p_N(\Delta)d\Delta$. (d) shows the same distribution on a linear scale for four different values of $N$.
		\label{distributions}}
\end{figure}

{\it Analytic derivations ---} We will next present analytic expressions for the central quantities governing the Poincar\'e revivals. We first demonstrate the exponential increase of the revival time $\tau_{\rm P}$ in the large $N$ limit for the system described above. Due to the central limit theorem, the probability distribution $p_N(\Delta)$ for $\Delta \equiv 1-p_e(\tau)$, will in this case follow the Gaussian distribution $p_N(\Delta)=\frac{1}{\sqrt{2\pi s^2}}e^{-(\Delta-\langle \Delta\rangle)^2/(2s^2)}$, where $s^2=\langle \Delta^2\rangle - \langle \Delta\rangle^2$ is the variance. For $N$ qubits with randomly distributed energies, we have $\langle \Delta\rangle =N \langle \Delta_j\rangle $ and $s^2 =N s_j^2$, where $\Delta_j$ and $s_j^2$ are the mean and the variance for individual qubits, respectively. For small threshold $\delta$ (i.e. near a recovery) we have 
\begin{equation}\label{gaussian-smallDelta}
	p_N(\Delta)\propto e^{-\langle \Delta\rangle^2/(2s^2)}\equiv e^{-bN},
\end{equation} 
where $b=\langle \Delta_j\rangle^2/(2s_j^2)$. The total probability $\mu_N(\delta)$ to stay below $\delta$ is then also $\propto e^{-bN}$. The Poincar\'{e} recovery time $\tau_{\rm P}$, i.e., the expectation value of the revival time $\tau_\delta$, on the other hand, is $\tau_{\rm P}\propto \mu_N^{-1}= e^{bN}$ as will be discussed below. 

For a more rigorous analytic derivation of $\mu_N$ and $\tau_{\rm P}$ (see details in the Supplemental Material), we focus on the dynamics of $\mathscr{C}_0(t)$, applying the given initial conditions and retaining only the first-order terms, yielding~\cite{Pekola2022}
\begin{equation}\label{1st-order-12}
	p_e(t)\equiv |\mathscr{C}_0(t)|^2= 
	1-4\sum_{j=1}^{N}{\mathcal{G}}_j^2\frac{\sin^2[(\lambda_j-1)\tau/2]}{(\lambda_j-1)^2},
\end{equation}
where ${\mathcal{G}}_j\equiv {g_{0j}}/{\hbar\Omega_0}$, $\lambda_j\equiv {\Omega_j}/{\Omega_0}$, and $\tau=\Omega_0 t$. We may identify $\Delta_j=4\mathcal{G}_j^2\sin^2(\phi_j/2)/\eta_j^2$, thus $p_e(\tau)\equiv 1-\Delta$, with $\Delta=\sum_{j=1}^{N}\Delta_j$. Here $\eta_j\equiv \lambda_j-1$ and $\eta_j\tau\equiv \phi_j$. For recoveries $p_e(\tau)\sim 1$, i.e., all the (positive) terms $\Delta_j$ in the sum are very small. The weak coupling approximation is well justified if $\mathcal{G}_j\lesssim 10^{-3}$, see Supplemental Material.
We realize that at sufficiently long times $\tau$, $\phi_j$ is randomly and uniformly distributed in the interval $[-\pi,\pi]$ modulo $2\pi$. For simplicity, we assume that the energies $\eta_j$ of the environment qubits are uniformly distributed around the central qubit in the interval $-\Delta\Omega/(2\Omega_0)<\eta_j<\Delta\Omega/(2\Omega_0)$. We then obtain the distribution of $\Delta_j$ for the case of equal coupling to all qubits, i.e. $\mathcal{G}_j\equiv \mathcal{G}$ for all $j=1,\,...\,,\,N$. To do this, we calculate the probability in the allowed range via $P(\Delta_j)=\int d\phi_j\int d\eta_j\, p(\phi_j)\,p(\eta_j)$. Here $p(\phi_j)$ and $p(\eta_j)$ are the uniform probability density functions. In the regime $\Delta_j<\Delta_{\rm m}=16\mathcal{G}^2/(\Delta\Omega/\Omega_0)^2$, we have
\begin{eqnarray}\label{full-prob-1stregion-m}
	P(\Delta_i)=\frac{2}{\pi}\bigg{[}\sqrt{\frac{\Delta_{\rm m}}{\Delta_j}-1}-\sqrt{\frac{\Delta_{\rm m}}{\Delta_j}}+\arcsin\big{(}\sqrt{\frac{\Delta_j}{\Delta_{\rm m}}}\big{)}\bigg{]}.
\end{eqnarray}
\begin{figure}
	\centering
	\includegraphics [width=\columnwidth] {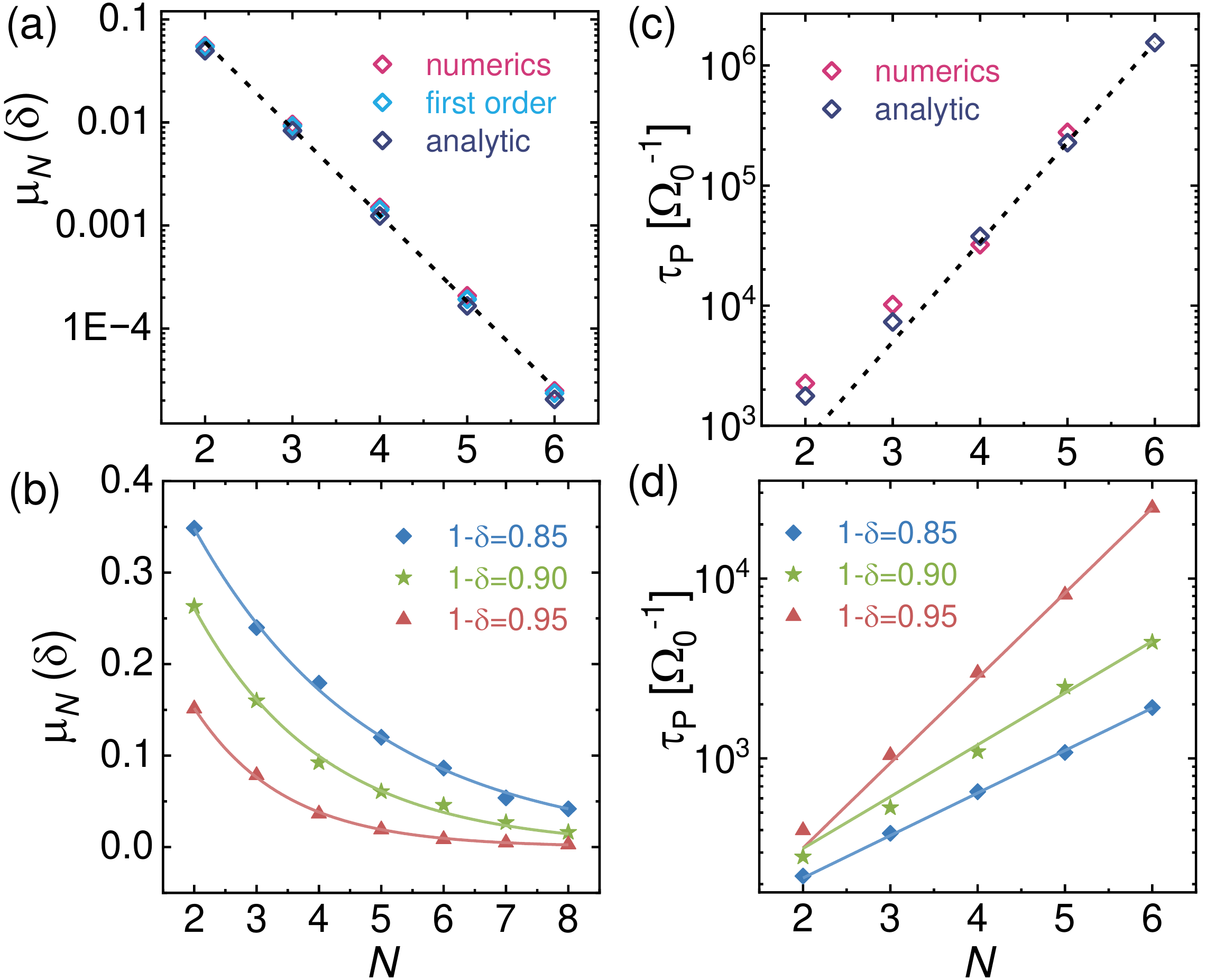}
	\caption{Revival probability $\mu_N(\delta)$ to have $p_e(\tau)>1-\delta$ in the long time limit, and the Poincar\'e time $\tau_{\rm P}$. (a) Comparison between the full numerical solution of the Schr\"odinger equation and the first-order approximation given by Eq.~\eqref{1st-order-12}, as well as the analytical solution given by Eq.~\eqref{final_prob_mu1}. The dashed line represents an exponential fit $\mu_N\propto e^{-1.93\,N}$. The parameters for this figure are $\Delta\Omega/\Omega_0=0.1$, $\mathcal{G}=0.001$, and $\delta=0.001$. (b) Numerical results for $\mu_N (\delta)$ for three values of $\delta$. In each case, $\mu_N(\delta)$ exhibits an exponential decay with $N$, with the corresponding fits $e^{-0.35N}$, $e^{-0.48N}$, and $e^{-0.69N}$, ordered from top to bottom and shown by solid lines. The parameters used in this panel are $\Delta\Omega/\Omega_0=0.1$ and $\Gamma_0=0.01\Omega_0$, where $\mathcal{G}_i$ uniformly distributed between $0$ and its maximum value determined by $\Gamma_0$. (c) Comparison of Poincare revival time $\tau_{\rm P}$ from the full numerical results and the analytical solution Eq.~\eqref {tau_P_v12}. Dashed line represents an exponential fit of the form $\tau_{\rm P} \propto \exp(bN)$ with $b=1.91$. Parameters are $\Delta\Omega/\Omega_0=0.1$, $\mathcal{G}=0.001$ and $\delta=0.001$. (d) Numerical results for the dependence of $\tau$ on $N$ for different threshold values of $\delta$. The data demonstrate the exponential increase predicted by the model, for $\Delta\Omega/\Omega_0=0.1$ and $\Gamma_0=0.01\Omega_0$, with uniform distributions of the couplings from $0$ to their maximum value determined by $\Gamma_0$.
		\label{muprobabilities}}
\end{figure}

The probability density is given by $p(\Delta_j)=dP(\Delta_j)/d\Delta_j$. For $\Delta_j\ll\Delta_{\rm m}$ we have 
\begin{equation}\label{prob-dens-1-m}
	p(\Delta_j)\simeq \frac{1}{\mathcal{T}}\frac{1}{\sqrt{\Delta_j}},
\end{equation}
with $\mathcal{T}=8\pi\mathcal{G}/(\Delta\Omega/\Omega_0)$. Equation~\eqref{prob-dens-1-m} allows us then to calculate the corresponding probability density for $\Delta$ as
\begin{equation}\label{final_pNDelta1}
	p_N(\Delta)=\frac{1}{(64\pi)^{\frac{N}{2}}\Gamma\big{(}\frac{N}{2}\big{)}}\bigg{(}\frac{\Delta\Omega/\Omega_0}{\mathcal{G}}\bigg{)}^N\,\Delta^{\frac{N}{2}-1}.
\end{equation}
Then the revival probability $\mu_N(\delta)=\int_{0}^{\delta} p_N(\Delta)d\Delta$ for being within the threshold, i.e., in the interval $1-p_e(\tau)<\delta$ is
\begin{equation}\label{final_prob_mu1}
	\mu_N(\delta)= \frac{1}{(64\pi)^{\frac{N}{2}}\Gamma\big{(}\frac{N}{2}+1\big{)}}\bigg{(}\frac{\Delta\Omega/\Omega_0}{\mathcal{G}}\bigg{)}^N\delta^{\frac{N}{2}}.
\end{equation}
The distributions $p_N(\Delta)$ and $\mu_N(\delta)$ for a few values of $N$ are shown in Fig.~\ref{distributions} together with the Gaussian approximation. This figure demonstrates the validity range of Eqs.~\eqref{final_pNDelta1} and \eqref{final_prob_mu1}, limited by the condition $\Delta_j\ll\Delta_{\rm m}$.
\begin{figure}
	\centering
	\includegraphics [width=\columnwidth] {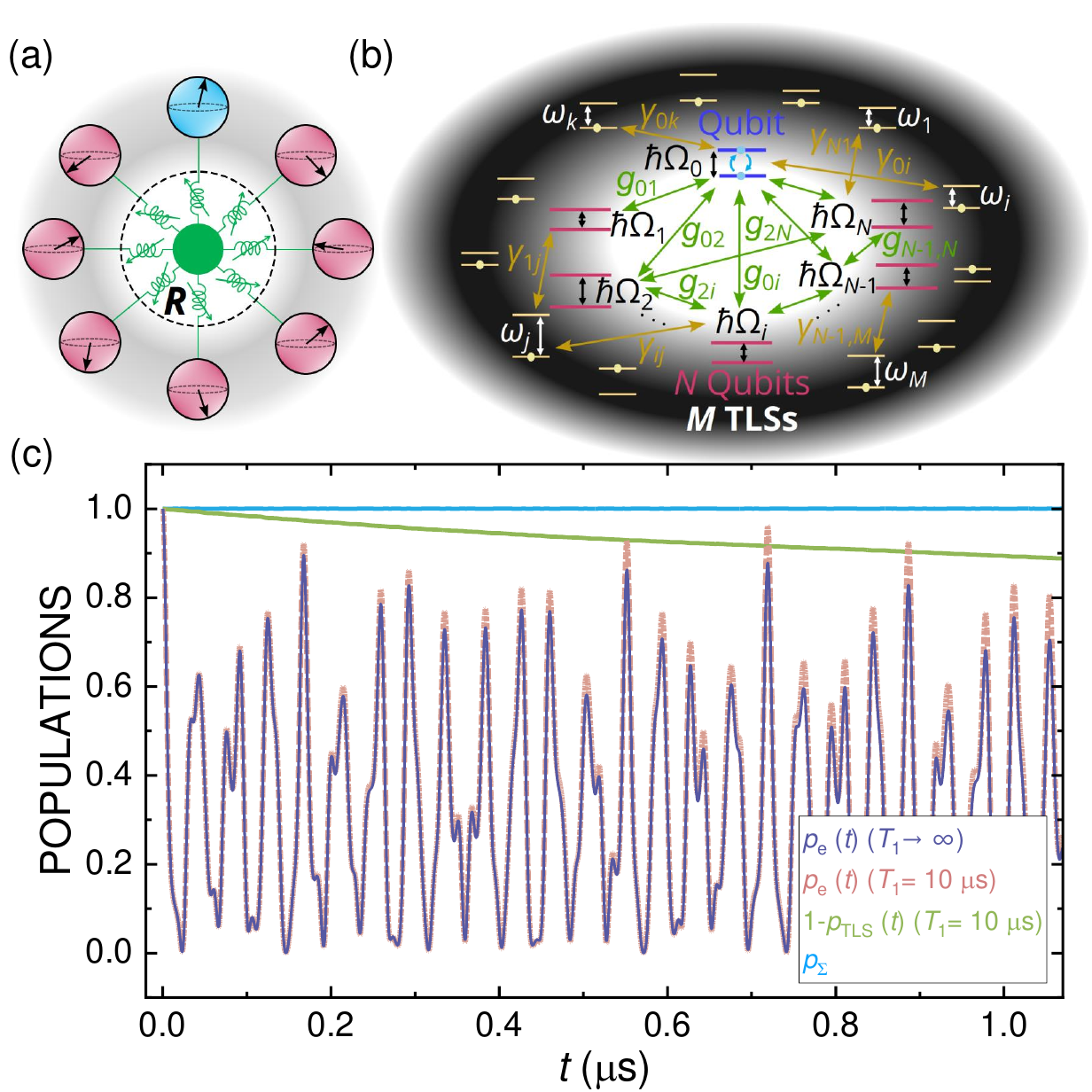}
	\caption{A possible experimental setup consists of $N+1$ qubits (one test qubit and $N$ environment qubits) coupled all-to-all via a superconducting router marked as $R$ within the dashed circle~\cite{Wu2024}. (a) The energies of all the qubits $\hbar\Omega_j$ and their couplings $g_{jk}$ are tunable. (b) A more realistic model of (a), taking into account the decoherence due to the external bath, which here is modeled by $M$ bath TLSs. (c) Numerical results for system in (b). We use a model size $N=5$, $M=10000$, $T_1=10~\mu_N$s, with the parameters $\Delta\Omega/\Omega_0=0.02$ and $\Gamma_0/\Omega_0=4\times 10^{-6}$. Purple and brown lines show the population $p_e(\tau)=|\mathscr{C}_0(t)|^2$ of the test qubit with and without relaxation, respectively. The green line is $\sum_{k=0}^{N}|\mathscr{C}_k(t)|^2$, showing the population in the $N+1$ test and environment qubits including relaxation, where decay results due to leakage to the TLS bath. Light blue line is $\sum_{k=0}^{N}|\mathscr{C}_k(t)|^2+\sum_{j=1}^{M}|D_j(t)|^2$, which remains at unity, demonstrating conservation of probability in the closed system formed by all the qubits and TLSs.
		\label{populations}}
\end{figure}

Next, we compare the analytical expression, Eq.~\eqref{final_prob_mu1}, with the full Schr\"odinger equation solution outlined in Section II and the outcome of Eq.~\eqref{1st-order-12}, both averaged over a long time interval. These results are shown in Fig.~\ref{muprobabilities}\,(a) for $N=2,\,3\,...,\,6$ for the parameters $\mathcal{G}=0.001$, $\Delta\Omega/\Omega_0=0.1$, and $\delta=0.001$, which are within the validity range of the analytic expression. We see that these results coincide in a satisfactory way and $\mu_N$ indeed decays exponentially, even for these small values of $N$. 

To evaluate the actual revival time $\tau_{\rm P}$, we find that the rate $r$ to cross the threshold at $p_e=1-\delta$ is given by 
\begin{equation}\label{r-1}
	r =  p_N(\delta)\langle R\rangle_+,
\end{equation}
where $\langle R\rangle_+\equiv \int_0^\infty dR\,p_R(R)R$
is the expectation value of $R\equiv -d\Delta/d\tau$ for positive slopes at $\Delta =\delta$ and $p_R(R)$ is the probability density of $R$. Incorporating the survival probability $P_s(\tau)$, we have ${dP_s(\tau)}/{d\tau}=-rP_s(\tau)$, which yields the revival time $\tau_{\rm P}=r^{-1}$ as
\begin{eqnarray}\label{tau_P_v12}
	\tau_{\rm P}=\frac{\sqrt{2}\pi\Gamma\big{(} \frac{N+1}{2}\big{)}}{N^{3/2} \Gamma\big{(} \frac{N}{2}\big{)}}\frac{\sqrt{\delta}}{\mathcal{G}}\mu_N(\delta)^{-1}. 
\end{eqnarray}

Panels~(c) and (d) in Fig.~\ref{muprobabilities} display the results for the Poincar\'e revival time $\tau_{\rm P}$ as a function of the environment size $N$. In panel \ref{muprobabilities}\,(c), we show the comparison of the numerical and analytic results, using the same parameters as in panel~\ref{muprobabilities}\,(a) for $\mu_N(\delta)$. We see an almost exponential increase of $\tau_{\rm P}$ with $N$, even for these low values of $N$, with good agreement between numerical and analytical results. The numerical results, outside the weak coupling regime, shown in panel~\ref{muprobabilities}\,(d), further demonstrate the exponential dependence on $N$, for three values of the revival threshold $\delta$.

{\it An experimentally realistic model ---} A recently presented setup~\cite{Wu2024} provides a superconducting multi-qubit platform that can be adapted to a possible test of the Poincar\'e revival, shown schematically in Fig.~\ref{populations}\,(a). To better match the actual experimental situation, we now include in our model the expected environmental decoherence, by introducing a bath of two-level systems (TLSs) surrounding and interacting with the $N+1$ test and environment qubits. The full Hamiltonian of the qubit system shown in Fig.~\ref{populations}\,(b) is given by
\begin{eqnarray}\label{Hamiltonian2}
	&&\mathcal{\hat{H}}=\hbar\Omega_0 \hat{a}^\dagger \hat{a}+\sum_{i=1}^{N} \hbar\Omega_i \hat{b}_i^\dagger \hat{b}_i+\sum_{j=1}^{M}\hbar\omega_j \hat{c}_j^\dagger \hat{c}_j+\nonumber\\&&\sum_{k=1}^{N} g_{0k}(\hat{a}^\dagger \hat{b}_k+\hat{b}_k^\dagger \hat{a})+\sum_{l\neq m}g_{lm}\hat{b}_l^\dagger \hat{b}_m+\sum_{n=1}^{M}\gamma_{0n}(\hat{a}^\dagger \hat{c}_n +\hat{c}_n^\dagger \hat{a})\nonumber\\&&+\sum_{p,q}\gamma_{pq}(\hat{b}_p^\dagger \hat{c}_q +\hat{c}_q^\dagger \hat{b}_p).
\end{eqnarray}
The first three terms describe the non-interacting Hamiltonian of the test qubit, the $N$ environment qubits, as previously, and the bath comprising $M$ two-level systems with energies $\hbar\omega_j$, respectively. The fourth and fifth terms represent the coupling between the test qubit and the environment qubits, as well as the interactions between the environment qubits. The final two terms describe the coupling of both the test qubit and the environment qubits to the $M$ TLSs, where the annihilation operators $\hat{c}_k$ are for the TLSs. In Eq.~\eqref{Hamiltonian2}, $g_{lm}$ denotes the coupling constant between the $l^{\rm th}$ and $m^{\rm th}$ environment qubit. Similarly, $\gamma_{0n}$ denotes the coupling constant between the central test qubit and the $n^{\rm th}$ TLS in the bath, while $\gamma_{pq}$ characterizes the coupling between the $p^{\rm th}$ environment qubit and the $q^{\rm th}$ bath TLS. We use a Fock state basis $|0\rangle\equiv |1,0\,0\,...\,0,\,0\,0\,...\,0\rangle$, $|{i}\rangle\equiv |0,0\,0\,...1 ({\rm i^{th}})\,...\,0,\,0\,0\,...\,0\rangle$, and $|\overline{j}\rangle\equiv |0,0\,0\,...\,0,\,0\,0\,...1 ({\rm j^{th}})\,...\,0\rangle$. Here, the basis vector is coded as follows: The first entry corresponds to the central qubit, the next $N$ entries correspond to each of the environment qubits, and the third group, running from $2+N$ to $2+N+M$, corresponds to the $M$ bath TLSs. These three sets are separated by commas in the ket vector for clarity. Based on this construction, the state $|\psi (t)\rangle$ can be written as
\begin{equation}\label{wavefunction}
	|\psi(t)\rangle=\mathscr{C}_0|0\rangle+\sum_{i=1}^{N}\mathscr{C}_i|{i}\rangle+\sum_{j=1}^{M} D_j|\overline{j}\rangle.
\end{equation}
In the interaction picture, the dynamics of the system are governed by the coupling terms via the perturbation $\mathbb{\hat{V}}_I(t)$, which arises from the last four terms of the Hamiltonian \eqref{Hamiltonian2}. We again solve the Schr\"odinger equation, $i\hbar \partial_t |\psi_I(t)\rangle=\mathbb{\hat{V}}_I(t)|\psi_I(t)\rangle$ for the complete system. This allows us to determine how the state of the entire system evolves over time (details given in SM). 

To assess the experimental feasibility, we develop a numerical experiment with parameters based on an existing system~\cite{Wu2024}. Taking into account the finite relaxation time of the qubits (here $T_1\sim 10~\mu_N$s), with realistic couplings and qubit energy distributions, we calculate the $p_e(\tau)$ pattern shown in Fig.~\ref{populations}\,(c). For reference, we also plot results for a fully isolated system (with no coupling to the TLS bath) with identical qubit parameters. Experiments lasting a few microseconds should not be significantly affected by qubit decoherence, but for precise analysis, one likely needs to renormalize the threshold values based on leakage to the external TLS bath. This is indicated by the decreasing probability for the excitation to remain within the $N+1$ qubit system, $\sum_{i=0}^{N}|\mathscr{C}_i(\tau)|^2$, as shown by the green line. 

%
{\it Conclusion ---} Whether and how unitary quantum systems thermalize remains a fundamental open question, with direct relevance for scaling quantum processors. A fingerprint of remaining quantum dynamics instead of thermalization is the revival of an initial state in a many-body system. Here, we have presented a quantitative analysis of the small $N$ behavior of the Poincar\'e recurrence in a quantum system and a blueprint for investigating these quantum revivals experimentally. The proposed construct uses a collection of coupled qubits that form a quasi-isolated quantum system. Our analysis also includes the decohering effects of a realistic environmental bath of TLSs, for which we derive explicit analytic and numerical results. Although our results are general, superconducting qubits are particularly suitable for an experimental study, as these systems can be tailored in terms of their energy spectrum and couplings with an exceptional degree of control. Furthermore, present-day superconducting qubits are sufficiently isolated from the environment, reducing the bath decoherence to an acceptable level. Most importantly, individual qubits can be measured accurately in the time domain, allowing the direct experimental study of Poincar\'e recurrence.

{\it Acknowledgments ---} We thank Paolo Muratore-Ginanneschi for many useful discussions. This work has received funding from the European Union’s Research and Innovation Programme, Horizon Europe, under the Marie Sklodowska-Curie Grant Agreement No. 101150440 (TcQTD). We acknowledge QuantERA II Programme that has received funding from the EU's H2020 research and innovation programme under the GA No 101017733, and the Research Council of Finland Centre of Excellence programme grant 336810 and grant 349601 (THEPOW). X. W. and A. N. C. acknowledge support from the Army Research Office and Laboratory for Physical Sciences (ARO W911NF2310077), the Air Force Office of Scientific Research (AFOSR FA9550-20-1-0270 and AFOSR MURI FA9550-23-1-0338), the NSF QLCI for HQAN (NSF award 2016136), the U.S. Department of Energy Office of Science National Quantum Information Science Research Center Q-NEXT, the Simons Foundation (Simons award 5099) and a 2024 Department of Defense Vannevar Bush Faculty Fellowship (ONR N000142512032). The authors declare no competing financial interests. \\
\clearpage
\onecolumngrid
\setcounter{section}{0}
\renewcommand{\thesection}{S\arabic{section}}
\setcounter{figure}{0} 
\renewcommand{\thefigure}{S\arabic{figure}} 
\renewcommand{\figurename}{Fig.} 
\setcounter{table}{0}
\renewcommand{\thetable}{S\arabic{table}}%
\setcounter{equation}{0} 
\renewcommand{\theequation}{S\arabic{equation}} 
\section*{Supplemental Material}

Here, we give the essential steps in deriving the mathematical expressions of the central quantities governing the revivals in the multi-qubit system.

\section{Analytic derivation of quantum recurrence properties in the linearly coupled qubit system}
The time dependent population, $p_e(\tau)$, of the central qubit in a single trajectory reads in the lowest order in coupling~\cite{Pekola2022}
\begin{equation}\label{1st-order-1}
	p_e(t)\equiv |\mathscr{C}_0(t)|^2=1-\frac{2}{\hbar^2}\sum_{j=1}^{N}{g}_{0j}^2\frac{1-\cos[(\Omega_j-\Omega_0)t]}{(\Omega_j-\Omega_0)^2}.
\end{equation}
Transforming to dimensionless quantities
\begin{equation}\label{0D-terms}
	{\mathcal{G}}_j\equiv {g_{0j}}/{\hbar\Omega_0},~\lambda_j\equiv {\Omega_j}/{\Omega_0},~{\rm and}~\tau=\Omega_0 t,
\end{equation}
Equation~\eqref{1st-order-1} reads
\begin{equation}\label{1st-order-2}
	p_e(\tau)=1-4\sum_{j=1}^{N}{\mathcal{G}}_j^2\frac{\sin^2[(\lambda_j-1)\tau/2]}{(\lambda_j-1)^2}\equiv 1-\sum_{j=1}^{N}\Delta_j= 1-\Delta.
\end{equation}
We are looking for recoveries, where $p_e(\tau)\sim 1$, i.e. all the positive terms in the sum are very small. 

\begin{figure}[h!]
	\centering
	\includegraphics [width=0.5\columnwidth] {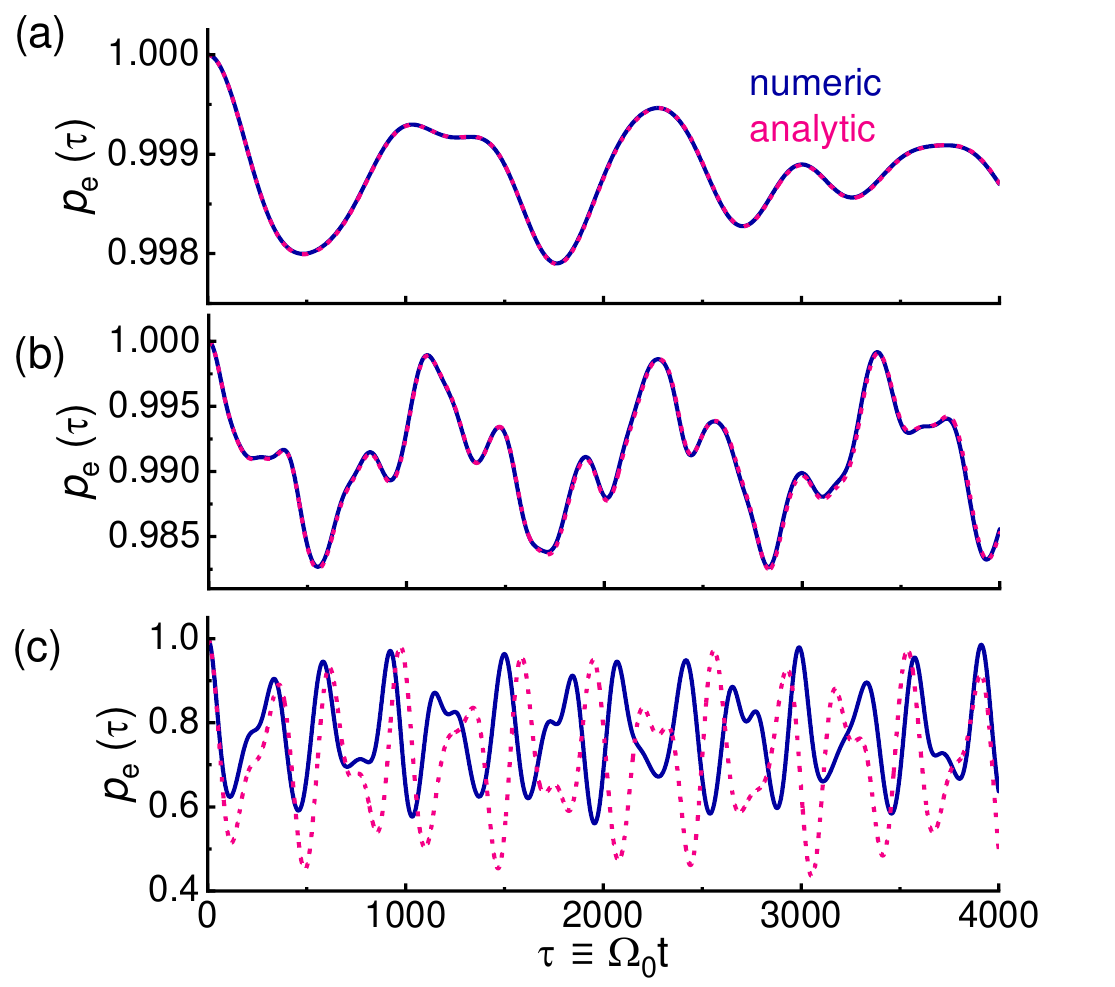}
	\caption{Time traces of the excited state population $p_e(\tau)$ of the central qubit.  (a)-(c) Comparison of the numerical solution of the Schr\"odinger equation and the weak coupling approximation of Eq.~\eqref{1st-order-1}. Here, $\mathcal{G}_j$ has a uniform distribution over the interval $0<\mathcal{G}_j<10^{-4}$ in (a), $0<\mathcal{G}_j<10^{-3}$ in (b), and $0<\mathcal{G}_j<10^{-2}$ in (c). Equation~\eqref  {1st-order-1} yields a good approximation for low values of $\mathcal{G}_j$, but fails when coupling gets stronger. In each case $p_e(\tau)$ represents an almost periodic function~\cite{Bohr1947,Amerio1971}, ideally in the longe-time limit.
		\label{Fig-traces_theoexp}}
\end{figure} 
A comparison between the numerical solution of the Schr\"odinger equation and the weak-coupling approximation from Eq.~\eqref {1st-order-1} is presented in Fig.~\ref {Fig-traces_theoexp} for three values of the coupling strength $\mathcal{G}_j$. As seen in panels (a) and (b), the approximation coincides with the numerical solution for small $\mathcal{G}_j$, whereas for larger coupling  it loses accuracy, as illustrated in panel (c).

\subsection{Distribution of $\Delta_j$}
\begin{figure}
	\centering
	\includegraphics [width=\columnwidth] {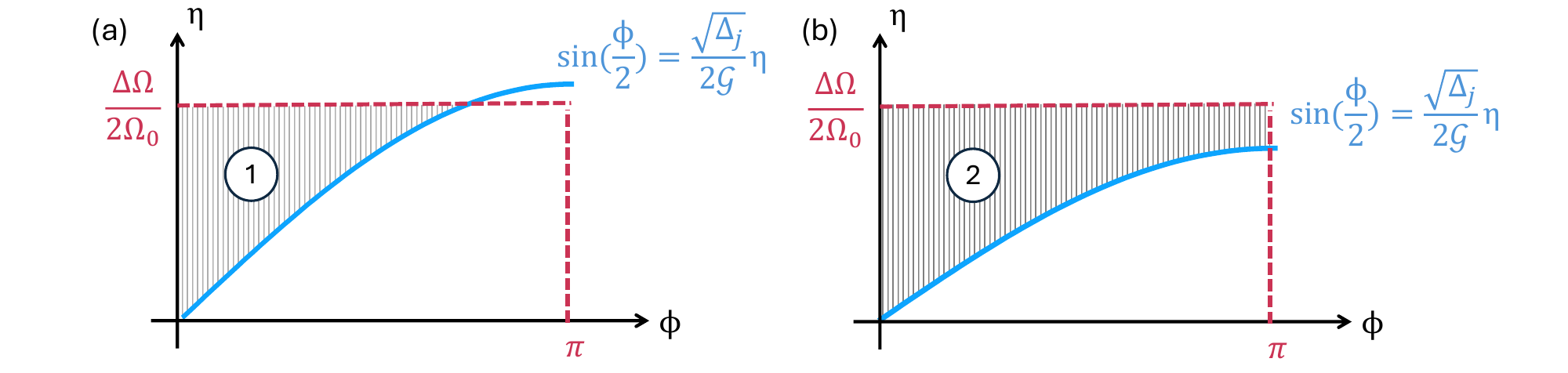}
	\caption{The shaded areas represent the regions where the integrals in Eqs.~\eqref{full-prob-1stregion} and \eqref{full-prob-2ndregion} have been evaluated, respectively.
		\label{integrals}}
\end{figure}

Our aim is to find the distribution of $\Delta$ for $N$ qubits, $p_N(\Delta)$, which serves as the generator of most analytic results in this paper. The analytical solution can be also written as
\begin{eqnarray}\label{1st-order-4}
	p_e(\tau)=1-4\sum_{j=1}^{N}\mathcal{G}_j^2\frac{\sin^2(\phi_j/2)}{\eta_j^2}\equiv 1-\sum_{j=1}^{N}\Delta_j.
\end{eqnarray}
Here, $\eta_j=\lambda_j-1$ and $\phi_j=\eta_j\tau$. For the rest of the analytic calculations we assume that the coupling $\mathcal{G}_j\equiv \mathcal{G}$ is equal for all qubits. We evaluate the full probability for $4\mathcal{G}^2\sin^2(\eta_j\tau/2)/\eta_j^2<\Delta_j$ as
\begin{equation}\label{full-prob-1}
	P(\Delta_j)=\int d\phi\int d\eta\, p(\phi)\,p(\eta),
\end{equation}
where we have $\phi\equiv\eta_j\tau/2$ and $\eta\equiv\eta_j$. Here $p(\phi)=\frac{1}{\pi}$ and $p(\eta)=\frac{2\Omega_0}{\Delta\Omega}$ are the corresponding probability densities in the regimes $0<\phi<\pi$ and $0< \eta<\Delta\Omega/(2\Omega_0)$. For the distribition of phase $\phi$ we have assumed that the time $\tau$ is long such that we can assume $\phi$ to be random in the range $0<\phi<\pi$ (in fact, the actual range is $[-\pi,\pi]$, but due to symmetry, we can focus only on $[0,\pi]$). Next, we solve the integral presented in Eq.~\eqref{full-prob-1} in the two regions shown in Fig.~\ref{integrals}. For the first and the important region where $\Delta_j<\Delta_{\rm m}=16\mathcal{G}^2/(\Delta\Omega/\Omega_0)^2$, we have
\begin{eqnarray}\label{full-prob-1stregion}
	P(\Delta_j)&&=\int_{0}^{\frac{\Delta\Omega}{2\Omega_0}}d\eta\int_{0}^{2\arcsin(\frac{\sqrt{\Delta_j}}{2\mathcal{G}}\eta)}d\phi\,\frac{1}{\pi}\frac{2\Omega_0}{\Delta\Omega}\nonumber\\&&=\frac{8}{\pi}\frac{\Omega_0}{\Delta\Omega}\frac{\mathcal{G}}{\sqrt{\Delta_j}}\bigg{[}\sqrt{1-\frac{\Delta_j}{16\mathcal{G}^2}\big{(}\frac{\Delta\Omega}{\Omega_0}\big{)}^2}-1+\frac{\sqrt{\Delta_j}}{4\mathcal{G}}\frac{\Delta\Omega}{\Omega_0}\arcsin\big{(}\frac{\sqrt{\Delta_j}}{4\mathcal{G}}\frac{\Delta\Omega}{\Omega_0}\big{)}\bigg{]}.
\end{eqnarray}
Substituting the expression for  $\Delta_{\rm m}$ we obtain Eq.~(4) in the main text. At the upper end of this region we have $P(\Delta_{\rm m})=1-2/\pi$. Using the full expression presented in Eq.~\eqref{full-prob-1stregion}, one can obtain the probability density given by $p(\Delta_j)=dP(\Delta_j)/d\Delta_j$. For $\Delta_j\ll\Delta_{\rm m}$ we have
\begin{eqnarray}\label{prob-dens-1}
	p(\Delta_j)&&=\frac{\Delta\Omega/\Omega_0}{8\pi\mathcal{G}}\frac{1}{\sqrt{\Delta_j}}\nonumber\\&&\equiv \frac{1}{\mathcal{T}}\frac{1}{\sqrt{\Delta_j}},
\end{eqnarray}
with $\mathcal{T}=\frac{8\pi\mathcal{G}}{\Delta\Omega/\Omega_0}$. For the second region, where $\Delta_j>\Delta_{\rm m}$, we have
\begin{eqnarray}\label{full-prob-2ndregion}
	P(\Delta_j)&&=\int_{0}^{\pi}d\phi\int_{\frac{2\mathcal{G}}{\sqrt{\Delta_j}}\sin(\frac{\phi}{2})}^{\frac{\Delta\Omega}{2\Omega_0}}d\eta\,\frac{1}{\pi}\frac{2\Omega_0}{\Delta\Omega}\nonumber\\&&=1-\frac{8}{\pi}\frac{\mathcal{G}}{\Delta\Omega/\Omega_0}\frac{1}{\sqrt{\Delta_j}}.
\end{eqnarray}
For this region we also have $P(\Delta_{\rm m})=1-2/\pi$, and the probability density at $\Delta_j>\Delta_{\rm m}$ reads
\begin{eqnarray}\label{prob-dens-2}
	p(\Delta_j)=\frac{4}{\pi}\frac{\mathcal{G}}{\Delta\Omega/\Omega_0}\Delta_j^{-3/2}.
\end{eqnarray}
Our discussion in the paper is focused on the first region $\Delta_j\ll \Delta_{\rm m}$ employing Eq.~\eqref{prob-dens-1}.

\subsection{Distribution of ${\Delta}$ close to revival $({\Delta\ll 1})$}

In order to obtain the probability density of $\Delta=\sum_{j=1}^{N}\Delta_j$, we use the characteristic function (the Fourier transform of the probability density function). For random variable $\Delta_j$ with a probability density $p(\Delta_j)$, its Fourier transform is
\begin{equation}\label{Fourierf-1}
	\mathcal{F}_j(\lambda)=\int_{-\infty}^{\infty} d\Delta_j\, e^{i\lambda\Delta_j}p(\Delta_j)\equiv \langle  e^{i\lambda\Delta_j}\rangle.
\end{equation}
We can then recover the probability density $p(\Delta_j)$ via the inverse Fourier transform as 
\begin{equation}\label{invFourierf-1}
	p(\Delta_j)=\frac{1}{2\pi}\int_{-\infty}^{\infty}d\lambda\,  e^{-i\lambda\Delta_j}\mathcal{F}_j(\lambda).
\end{equation}
For $\Delta$ we have the corresponding characteristic function $\mathcal{F}(\lambda)=\langle e^{-i\lambda\sum_{j=1}^{N}\Delta_j}\rangle$. In order to determine the complete probability density, we again write
\begin{equation}\label{gen-PDF-1}
	p_N(\Delta)=\frac{1}{2\pi}\int_{-\infty}^{\infty}d\lambda e^{-i\lambda \Delta}\mathcal{F}(\lambda)
\end{equation}
We begin by writing the expression for $\mathcal{F}(\lambda)$ with the help of Eq.~\eqref{prob-dens-1} in the small $\Delta_j$ regime as
\begin{eqnarray}\label{gen-PDF-1-3}
	\mathcal{F}(\lambda)&&=\langle e^{i\lambda \sum_{j=1}^{N}\Delta_j} \rangle=\int_{0}^{\Delta}d\Delta_1\int_{0}^{\Delta-\Delta_1}d\Delta_2...\int_{0}^{\Delta-\Delta_1-\Delta_2-...-\Delta_{N-1}}d\Delta_N\, e^{i\lambda \Delta_1} e^{i\lambda \Delta_2}... e^{i\lambda \Delta_N} p(\Delta_1)p(\Delta_2)...p(\Delta_N)\nonumber\\&&=\frac{1}{\mathcal{T}^N}\int_{0}^{\Delta}d\Delta_1\int_{0}^{\Delta-\Delta_1}d\Delta_2...\int_{0}^{\Delta-\Delta_1-\Delta_2-...-\Delta_{N-1}}d\Delta_N \frac{ e^{i\lambda \sum_{j=1}^{N}\Delta_j}}{\sqrt{\Delta_1\Delta_2...\Delta_N}}.
\end{eqnarray}
Substituting Eq.~\eqref{gen-PDF-1-3} in Eq.~\eqref{gen-PDF-1}, we have
\begin{eqnarray}\label{gen-PDF-2-2}
	p_N(\Delta)&&=\frac{1}{\mathcal{T}^N}\int_{-\infty}^{\infty}\frac{d\lambda}{2\pi}e^{-i\lambda \Delta}\int_{0}^{\Delta}d\Delta_1\int_{0}^{\Delta-\Delta_1}d\Delta_2...\int_{0}^{\Delta-\Delta_1-\Delta_2-...-\Delta_{N-1}}d\Delta_N \frac{ e^{i\lambda \sum_{j=1}^{N}\Delta_j}}{\sqrt{\Delta_1\Delta_2...\Delta_N}}
	\nonumber\\&&=\frac{1}{\mathcal{T}^N}\int_{0}^{\Delta}\frac{d\Delta_1}{\sqrt{\Delta_1}}\int_{0}^{\Delta-\Delta_1}\frac{d\Delta_2}{\sqrt{\Delta_2}}...\int_{0}^{\Delta-\Delta_1-\Delta_2-...-\Delta_{N-1}}\frac{d\Delta_N}{\sqrt{\Delta_N}}\int_{-\infty}^{\infty}\frac{d\lambda}{2\pi}e^{i\lambda (\sum_{j=1}^{N}\Delta_j-\Delta)}. 
\end{eqnarray}
The solution for the last integral reads $\int_{-\infty}^{\infty}\frac{d\lambda}{2\pi}e^{i\lambda (\sum_{j=1}^{N}\Delta_j-\Delta)}=\delta(\sum_{j=1}^{N}\Delta_j-\Delta)$; here, $\delta(x)$ is the Dirac delta function. Then we have
\begin{equation}\label{gen-PDF-3}
	p_N(\Delta)=\frac{1}{\mathcal{T}^N}\int_{0}^{\Delta}\frac{d\Delta_1}{\sqrt{\Delta_1}}\int_{0}^{\Delta-\Delta_1}\frac{d\Delta_2}{\sqrt{\Delta_2}}...\int_{0}^{\Delta-\Delta_1-\Delta_2-...-\Delta_{N-2}}\frac{d\Delta_{N-1}}{\sqrt{\Delta_{N-1}}}\frac{1}{\sqrt{\Delta-\sum_{j=1}^{N-1}\Delta_j}}.
\end{equation}
Equation~\eqref{gen-PDF-3} becomes
\begin{equation}\label{gen-PDF-4}
	p_N(\Delta)=\frac{1}{\mathcal{T}^N}a_N\Delta^{\frac{N}{2}-1},
\end{equation}
where 
\begin{equation}\label{gen-PDF-4-coe/an}
	a_N=\int_{0}^{1}\frac{du_1}{\sqrt{u_1}}\int_{0}^{1-u_1}\frac{du_2}{\sqrt{u_2}}...\int_{0}^{1-u_1-...-u_{N-2}}\frac{du_{N-1}}{\sqrt{u_{N-1}(1-\sum_{j=1}^{N-1}u_j)}}=\frac{\pi^\frac{N}{2}}{\Gamma(\frac{N}{2})}.
\end{equation}
This yields our final result for small $\Delta$
\begin{equation}\label{final_pNDelta}
	p_N(\Delta)=\frac{1}{(64\pi)^{\frac{N}{2}}\Gamma\big{(}\frac{N}{2}\big{)}}\bigg{(}\frac{\Delta\Omega/\Omega_0}{\mathcal{G}}\bigg{)}^N\,\Delta^{\frac{N}{2}-1}.
\end{equation}
Then the probability of finding the qubit in the range $p_e(\tau)>1-\delta$ is given by
\begin{eqnarray}\label{totalprob-1}
	\mu(\delta)=P(\Delta<\delta)~&&=\int_{0}^{\delta}p_N(\Delta)d\Delta=\frac{2\delta}{N}p_N(\delta)\nonumber\\&&=\frac{1}{(64\pi)^{\frac{N}{2}}\Gamma\big{(}\frac{N}{2}+1\big{)}}\bigg{(}\frac{\Delta\Omega/\Omega_0}{\mathcal{G}}\bigg{)}^N\delta^{\frac{N}{2}}.
\end{eqnarray}

\subsection{Derivation of Poincar\'e revival time ${\tau_{\rm P}}$} 
The rate $r$ to cross the threshold at $p_e=1-\delta$ is given by 
\begin{equation}
	r =P(p_e(\tau+\delta \tau) > 1-\delta\, |\, p_e(\tau) < 1-\delta)/\delta \tau,
\end{equation}
where ``$|$" refers to conditional probability and $\delta\tau$ is a small (eventually infinitesimal) time interval. 

If we then denote $R\equiv -d\Delta/d\tau$, i.e., the rate of change of $p_e(\tau)$, we can write the probability to cross the threshold within time $\delta\tau$ as $\int dR\int d\Delta p_R(R) p_N(\Delta)$, under the conditions $R>0$ and $\delta<\Delta<\delta+R\delta\tau$. Here $p_R(R)$ is the probability distribution of $R$ in the vicinity of $\delta$. Then the rate can be recast into the form
\begin{equation}
	r = \frac{1}{\delta \tau}\int_0^\infty dRp_R(R)\int_\delta^{\delta+R\delta \tau}d\Delta p_N(\Delta).
\end{equation}
For $\delta\tau \rightarrow 0$, we find 
\begin{equation}
	r =  p_N(\delta)\langle R\rangle_+,
\end{equation}
where $\langle R\rangle_+=\int_0^\infty dRp_R(R)R$
is the expectation value of $R$ for positive slopes at $\Delta=\delta$. Next, we evaluate $\langle R\rangle_+$. To proceed, we start with $\Delta_j=4\mathcal{G}^2\sin^2(\eta_j\tau/2)/\eta_j^2$, and we have $R_j=4\mathcal{G}^2\sin(\eta_j\tau/2)\cos(\eta_j\tau/2)/\eta_j$. Since the probability of crossing the threshold in either direction is equal, we approximate $R_j$  as $R_j=2\mathcal{G}\sigma_j\sqrt{\Delta_j}$, where $\sigma_j=\pm 1$ represents the random and equally probable values of $\cos(\eta_j\tau/2)$ at $\eta_j\tau/2=0,\,\pi,\,2\pi\,,...$. Next, we calculate $\langle R\rangle_+$, where we have $\langle R\rangle_+=\langle \sum_{j=1}^{N} R_j\rangle_+=\langle 2\mathcal{G}\sum_{j=1}^{N} \sigma_j\sqrt{\Delta_j}\rangle_{+,\delta}$. Here, the subscript $+$ represents the positive direction and we calculate the expectation at the threshold value $\delta$. Since $\sigma_j$ and $\sqrt{\Delta_j}$ (for a random time instant, each qubit can have either a positive or negative slope with equal probability), are uncorrelated we can write $\langle  R\rangle_+=2\mathcal{G} \langle \sum_{j=1}^{N} \sigma_j\rangle_+ \langle \sqrt{\Delta_j}\rangle_\delta$.

First, we evaluate $\langle \sum_{j=1}^{N} \sigma_j\rangle_+$. The quantity $\sigma_j=\pm 1$ follows a binomial distribution. By considering only the positive contributions corresponding to the rising $\delta$, we have for small number of $N$
\begin{eqnarray}\label{bionomial-smallN-1}
	\langle \sum_{j=1}^{N} \sigma_j\rangle_+&&=\frac{N\,2^{-(N+1)}N!}{\big{[}\big{(}\frac{N}{2}\big{)}!\big{]}^2},~~~~~N=2,4,6,...\nonumber\\&&=\frac{N\,2^{-N}(N-1)!}{\big{[}\big{(}\frac{N-1}{2}\big{)}!\big{]}^2},~~N=3,5,7,...
\end{eqnarray}
In the large-$N$ limit, the distribution of $\sigma\equiv \sum_{j=1}^{N}\sigma_j$ approches a Gaussian. In this case we have
\begin{eqnarray}\label{gaussian-sigma-1}
	&&\langle \sigma\rangle=\sum_{j=0}^{N}(2j-N)\binom{N}{j} (\frac{1}{2})^N=0\nonumber\\&&\langle \sigma^2\rangle=\sum_{j=0}^{N}(2j-N)^2\binom{N}{j} (\frac{1}{2})^N=N.
\end{eqnarray}
Then the variance will be $\langle \delta\sigma^2\rangle=\langle \sigma^2\rangle-\langle \sigma\rangle^2=N$. In this case the Gaussian approximation is given by $p(\sigma)=\frac{1}{\sqrt{2\pi N}}\exp(-\sigma^2/(2N))$. From this we can write:
\begin{equation}\label{gaussian-largeN-1}
	\langle \sigma\rangle_+=\int_{0}^{\infty}d\sigma\,\sigma\,p(\sigma)=\sqrt{\frac{N}{2\pi}}.
\end{equation}
\begin{figure}
	\centering
	\includegraphics [width=0.5\columnwidth] {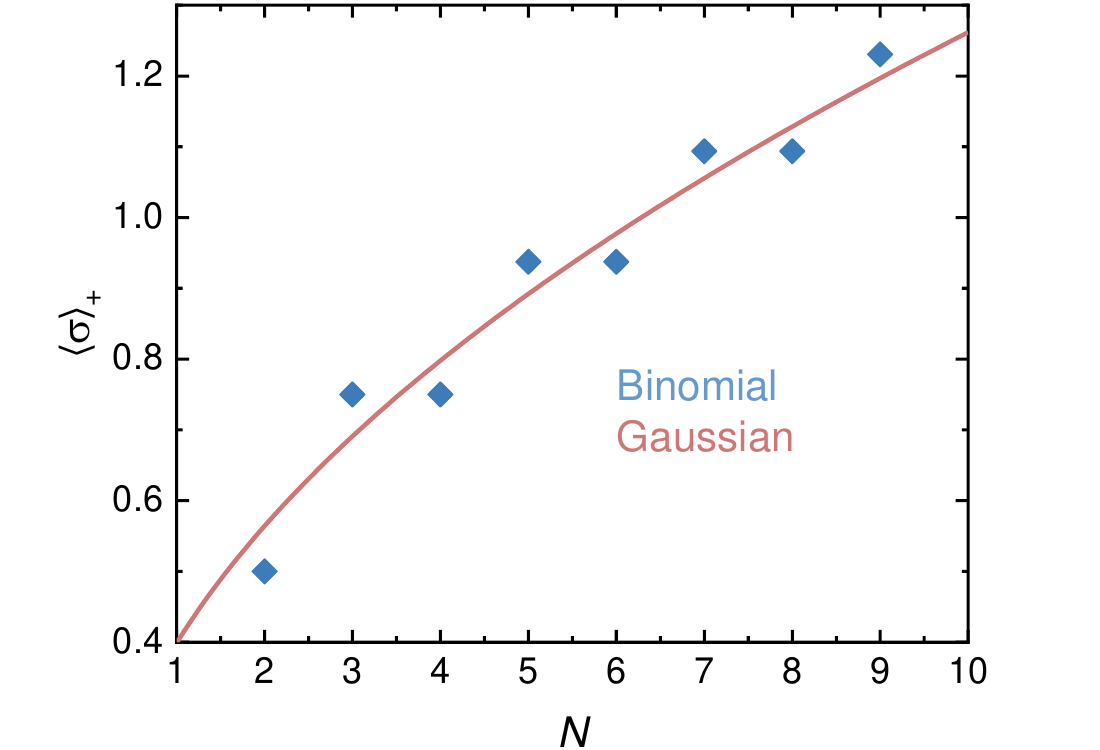}
	\caption{Binomial vs Gaussian approximation for $\langle \sigma \rangle_+\equiv\langle \sum_{j=1}^N \sigma_j \rangle_+$.
		\label{binomialvsgaussian}}
\end{figure}
Figure~\ref {binomialvsgaussian} shows a comparison between the binomial values and its Gaussian approximation for  $\langle \sum_{j=1}^N \sigma_j \rangle$ for different values of $N$.

The expectation value of $\sqrt{\Delta_j}$, is given by $\langle \sqrt{\Delta_j}\rangle=\int_{0}^{\delta}d\Delta_j\,\sqrt{\Delta_j}\,p^{(\delta)}(\Delta_j)$ where $p^{(\delta)}(\Delta_j)$ is given by
\begin{equation}\label{p_delta_v1}
	p^{(\delta)}(\Delta_j)=\frac{p_1(\Delta_j)p_{N-1}(\delta-\Delta_j)}{{\int_{0}^{\delta}d\Delta_j}p_1(\Delta_j)p_{N-1}(\delta-\Delta_j)}.
\end{equation}
Here the denominator equals $p_N(\Delta)$. Using Eq.~\eqref{gen-PDF-4} we then have
\begin{equation}\label{p_delta_v2}
	p^{(\delta)}(\Delta_j)=\frac{\frac{1}{\sqrt{\Delta_j}}a_{N-1}(\delta-\Delta_j)^{\frac{N}{2}-\frac{3}{2}}}{a_N\delta^{\frac{N}{2}-1}}.
\end{equation}
Then $\langle \sqrt{\Delta_j}\rangle$ is given by
\begin{equation}\label{Delta_ave_v1}
	\langle \sqrt{\Delta_j}\rangle=\frac{\int_{0}^{\delta}d\Delta_j\,a_{N-1}(\delta-\Delta_j)^{\frac{N}{2}-\frac{3}{2}}}{a_N\delta^{\frac{N}{2}-1}}.
\end{equation}
An elementary integration yields
\begin{eqnarray}\label{Delta_ave_v2}
	\langle \sqrt{\Delta_j}\rangle=\frac{2a_{N-1}/a_N}{N-1}\delta=\frac{1}{\sqrt{\pi}}\frac{\Gamma\big{(}\frac{N}{2}\big{)}}{\Gamma\big{(}\frac{N+1}{2}\big{)}}\sqrt{\delta}.
\end{eqnarray}
This equation indicates that $\langle \sqrt{\Delta_j}\rangle$ is identical for all $j$.

Using Eqs.~\eqref{gaussian-largeN-1} and \eqref{Delta_ave_v2}, we then have for $\langle R\rangle_+$ 
\begin{equation}\label{Rave+_1}
	\langle R\rangle_+=\frac{\sqrt{2N}}{\pi}\frac{\Gamma\big{(}\frac{N}{2}\big{)}}{\Gamma\big{(}\frac{N+1}{2}\big{)}} \mathcal{G}\sqrt{\delta}.
\end{equation} 
As a final step, incorporating the survival probability $P_s(\tau)$, we have $\frac{dP_s(\tau)}{d\tau}=-rP_s(\tau)$, which yields
\begin{eqnarray}\label{tau_P_v1}
	\tau_{\rm P}~&&=-\int_{0}^{\infty}d\tau\,\tau\,\frac{dP_s(\tau)}{d\tau}=\frac{1}{r}\nonumber\\&&=\frac{\sqrt{2}\pi\Gamma\big{(} \frac{N+1}{2}\big{)}}{N^{3/2} \Gamma\big{(} \frac{N}{2}\big{)}}\frac{\sqrt{\delta}}{\mathcal{G}}\mu(\delta)^{-1}.
\end{eqnarray}

\section{An emsemble of qubits coupled to a bath of two-level systems}
\begin{figure}[h!]
	\centering
	\includegraphics [width=0.65\columnwidth] {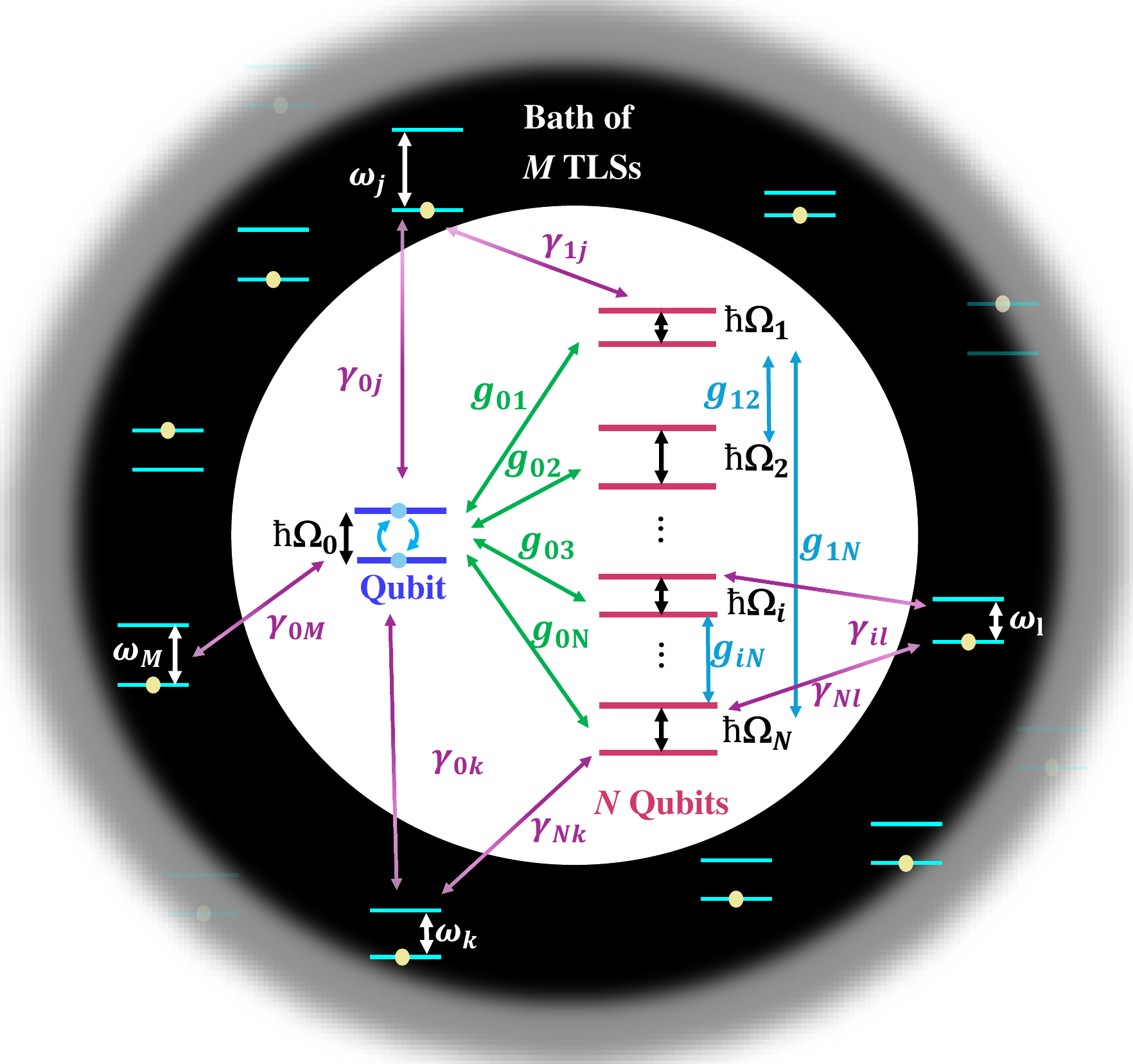}
	\caption{The setup; mimicking the experimental configuration. 
		\label{setup}}
\end{figure}
To assess an experimental platform~\cite{Wu2024}, we analyze the previous system embedded within a bath. We write the total Hamiltonian of the setup shown in Fig.~\ref{setup} as
\begin{equation}\label{Hamiltonian}
	\mathcal{\hat{H}}=\hbar\Omega_0 \hat{a}^\dagger \hat{a}+\sum_{i=1}^{N} \hbar\Omega_i \hat{b}_i^\dagger \hat{b}_i+\sum_{j=1}^{M}\hbar\omega_j \hat{c}_j^\dagger \hat{c}_j+\sum_{k=1}^{N} g_{0k}(\hat{a}^\dagger \hat{b}_k +\hat{b}_k^\dagger \hat{a})+\sum_{l\neq m}g_{lm}\hat{b}_l^\dagger \hat{b}_m+\sum_{n=1}^{M}\gamma_{0n}(\hat{a}^\dagger \hat{c}_n +\hat{c}_n^\dagger \hat{a})+\sum_{p,q}\gamma_{pq}(\hat{b}_p^\dagger \hat{c}_q +\hat{c}_q^\dagger \hat{b}_p).
\end{equation}
The first three terms represent the noninteracting Hamiltonian of the central qubit, the $N$ environment qubits, and the bath, formed of $M$ two-level systems (TLSs), respectively. The fourth to sixth terms correspond to the coupling Hamiltonian between the central qubit to other qubits, the interaction among the $N$ other qubits, and finally, the coupling of all qubits with the TLSs in the environment. The ladder operators in the expression $\hat{a}^\dagger(\hat{a})$, $\hat{b}_j^\dagger(\hat{b}_j)$, and $\hat{c}_k^\dagger(\hat{c}_k)$ are the creation (annihilation) operators for the central qubit, environment qubits, and TLSs, respectively. In \eqref{Hamiltonian}, $g_{0i}$ denotes the coupling constant between the central qubit (with energy $\hbar\Omega_0$) and the $i^{\rm th}$ environment qubit (with energy $\hbar\Omega_i$). Similarly, $g_{ik}$ represents the coupling constant between the $i^{\rm th}$ and $k^{\rm th}$ environment qubit, while $\gamma_{jk}$ describes the coupling between the $j^{\rm th}$ qubit and the $k^{\rm th}$ TLS in the bath, which has energy $\hbar\omega_k$.

The basis of the Fock states that we use is formed of $|0\rangle\equiv |1,0~0~...~0,~0~0~...~0\rangle$, $|\overline{i}\rangle\equiv |0,0~0~...1 ({\rm i^{th}})~...~0,~0~0~...~0\rangle$, and $|\underline{j}\rangle\equiv |0,0~0~...~0,~0~0~...1 ({\rm j^{th}})~...~0\rangle$. Here, the basis vector is structured as follows: the first entrance corresponds to the central qubit, the second to the $N+1^{^{\rm th}}$ entries correspond to each of the $N$ qubits, and the third group, spanning from $2+N$ to $2+N+M$, corresponds to the $M$ TLSs. These three sets are separated by commas for clarity. The state $|\psi (t)\rangle$ is given in this basis by
\begin{equation}\label{wavefunction}
	|\psi(t)\rangle=\mathscr{C}_0|0\rangle+\sum_{i=1}^{N}\mathscr{C}_i|\overline{i}\rangle+\sum_{j=1}^{M} D_j|\underline{j}\rangle.
\end{equation}

The last three terms of Eq.~\eqref{Hamiltonian} are given in the interaction picture by
\begin{equation}\label{interactionpic}
	\mathbb{\hat{V}}_I(t)=\sum_{k=1}^{N} g_{0k}(\hat{a}^\dagger \hat{b}_k e^{i(\Omega_0-\Omega_k)t}+\hat{b}_k^\dagger \hat{a}e^{-i(\Omega_0-\Omega_k)t})+\sum_{l\neq m}g_{lm}\hat{b}_l^\dagger \hat{b}_m e^{i(\Omega_l-\Omega_m)t})+\sum_{p,q}\gamma_{pq}(\hat{b}_p^\dagger \hat{c}_q e^{i(\Omega_p-\omega_q)t} +\hat{c}_q^\dagger \hat{b}_p e^{-i(\Omega_p-\omega_q)t}).
\end{equation}
Our goal is to solve the Schr\"odinger equation $i\hbar \partial_t |\psi_I(t)\rangle=\mathbb{\hat{V}}_I(t)|\psi_I(t)\rangle$ for the whole system. With the help of Eqs.~\eqref{wavefunction} and \eqref{interactionpic} we have
\begin{eqnarray}\label{schrodinger-1}
	i\hbar \bigg{(}\dot{\mathscr{C}}_0|0\rangle+\sum_{i=1}^{N}\dot{\mathscr{C}}_i|\overline{i}\rangle+\sum_{j=1}^{M}\dot{D}_j|\underline{j}\rangle\bigg{)}&&=\mathscr{C}_0\sum_{k=1}^{N} g_{0k}e^{-i(\Omega_0-\Omega_k)t}|\overline{k}\rangle+\sum_{i=1}^{N}\mathscr{C}_i g_{0i}e^{i(\Omega_0-\Omega_i)t}|0\rangle+\sum_{l\neq i}\mathscr{C}_i g_{li}e^{i(\Omega_l-\Omega_i)t}|\overline{l}\rangle\nonumber\\&&+\sum_{i,q}\mathscr{C}_i\gamma_{iq}e^{-i(\Omega_i-\omega_q)t}|\underline{q}\rangle+\sum_{p,j}D_j\gamma_{pj}e^{i(\Omega_p-\omega_j)t}|\overline{q}\rangle.
\end{eqnarray}
Comparing the two sides of the above equation, the time evolution of the state of the system is given by
\begin{eqnarray}\label{set-of-equations-1}
	&&\dot{\mathscr{C}}_0=-\frac{i}{\hbar}\sum_{i=1}^{N}\mathscr{C}_i\, g_{0i}\,e^{i(\Omega_0-\Omega_i)t}-\frac{i}{\hbar}\sum_{k=1}^{M}D_k\, \gamma_{0k}\,e^{i(\Omega_0-\omega_k)t}\nonumber\\&&\dot{\mathscr{C}}_i=-\frac{i}{\hbar}\mathscr{C}_0\, g_{0i}\,e^{-i(\Omega_0-\Omega_i)t}-\frac{i}{\hbar}\sum_{j=1}^{N}\mathscr{C}_j\, g_{ij}\,e^{i(\Omega_i-\Omega_j)t}-\frac{i}{\hbar}\sum_{k=1}^{M}D_k\, \gamma_{ik}\,e^{i(\Omega_i-\omega_k)t}\nonumber\\&&\dot{D}_j=-\frac{i}{\hbar}\mathscr{C}_0\, \gamma_{0j}\,e^{-i(\Omega_0-\omega_j)t}-\frac{i}{\hbar}\sum_{l=1}^{M}\mathscr{C}_l\,\gamma_{lj}\,e^{-i(\Omega_l-\omega_j)t}.
\end{eqnarray}
{\bf The argument about dephasing:} Dephasing arises from (slow) variation of qubit energy $\Omega_j (\tau)=\Omega_j+\delta \Omega_j (\tau)$~\cite{pekola2015}. The Hamiltonian then reads
\begin{equation}\label{Hamilton-dephase-1}
	\mathcal{\hat{H}}=\hbar (\Omega_0+\delta \Omega (\tau))\hat{a}^\dagger \hat{a}+\sum \hbar(\Omega_j+\delta \Omega_j (\tau)) \hat{b}_i^\dagger \hat{b}_i +\mathbb{\hat{V}}(\tau),
\end{equation}
where $\mathbb{\hat{V}}(\tau)$ is the coupling presented above. We may assume that for the short time of the measurement, the energies of the qubits are slightly adjusted by $\delta \Omega_j(0)$ for $j=0\,1,...,N+1$ from the set values. If this noise is small ($\delta\Omega(0)$, $\delta \Omega_j(0) \ll \Delta \Omega$ for all $j$:s), we may ignore its influence on the results.

\bibliography{ref}

	\end{document}